\documentclass[12pt]{article}
\usepackage{amssymb}
\usepackage{graphicx}
\usepackage{psfrag}
\usepackage{xcolor}

\begin{document}

\begin{center}

{\Large \bf Massive evaluation and analysis of Poincar\'e recurrences on
grids of initial data: a tool to map chaotic diffusion}

\bigskip

Ivan I. Shevchenko$^{a,b}$\footnote{E-mail:
ivan.i.shevchenko@gmail.com}, Guillaume Rollin$^c$,
Alexander~V.~Melnikov$^d$, Jos\'e~Lages$^c$

\medskip

$^a$~{\it Institute of Applied Astronomy, RAS, 191187 Saint Petersburg, Russia } \\
$^b$~{\it Lebedev Physical Institute, RAS, 119991 Moscow, Russia } \\
$^c$~{\it Institut UTINAM, OSU THETA, CNRS,
Universit\'e de Bourgogne Franche-Comt\'e, Besan\c{c}on 25030, France } \\
$^d$~{\it Tomsk State University, 634050 Tomsk, Russia }

\end{center}

\begin{abstract}
We present a novel numerical method aimed to characterize global
behaviour, in particular chaotic diffusion, in dynamical systems.
It is based on an analysis of the Poincar\'e recurrence statistics
on massive grids of initial data or values of parameters. We
concentrate on Hamiltonian systems, featuring the method
separately for the cases of bounded and non-bounded phase spaces.
The embodiments of the method in each of the cases are specific.
We compare the performances of the proposed \textit{Poincar\'e
recurrence method} (PRM) and the custom \textit{Lyapunov exponent}
(LE) methods and show that they expose the global dynamics almost
identically. However, a major advantage of the new method over the
known global numerical tools, such as LE, FLI, MEGNO, and FA, is
that it allows one to construct, in some approximation, charts of
local diffusion timescales. Moreover, it is algorithmically simple
and straightforward to apply.
\end{abstract}



\newpage

\section*{Introduction}
\label{intro}

A number of numerical tools, such as based on computation of
Lyapunov exponents (LE), fast Lyapunov indicators (FLI), mean
exponential growth number (MEGNO), and frequency analysis (FA),
have been elaborated up to now to explore dynamical systems in
global contexts (for a review, see, e.g., \cite{Morbi02}).
However, none of the known global tools allows one to expose
diffusion rates globally. To elaborate a global numerical tool
that overcomes this difficulty is just the aim of the present
study. Therefore, we propose and develop a novel general method,
based on a massive numerical analysis of Poincar\'e recurrences of
orbits on fine grids of initial data or values of parameters.
What makes the new method complementary to (and often advantageous
over) other global numerical tools, such as LE, FLI, MEGNO, and
FA, is that it allows one to characterize local diffusion rates.
Indeed, LE, FLI, and MEGNO characterize local divergence of
trajectories, and FA their spectral properties. Therefore, the
output of the other methods does not have any universal relation
to diffusion rates, whereas these are just the diffusion rates
that are often needed.

The paper is organized as follows. In Section~1, we briefly review
the known tools aimed to study global dynamics (LE, FLI, MEGNO,
FA). In Section~2, concentrating on Hamiltonian systems, we
introduce a novel global numerical method, the \textit{Poincar\'e
recurrence method} (PRM). We feature this new method separately
for the cases of bounded and non-bounded phase spaces, and show
that the embodiments of the method in each of the cases are
specific. We compare the performances of the PRM and LE methods
and show that they expose the global dynamics almost identically,
but PRM is algorithmically much simpler, and it is straightforward
to apply. In Section~3, we concentrate on theoretical issues of
the Poincar\'e recurrence statistics and identify the major
advantage of the new method over the custom global numerical
tools: by providing the opportunity to construct charts of
diffusion rates, it allows one to assess the timescales of
\textit{clearing} of chaotic domains of phase space in various
physical and astrophysical applications. Section~4 is devoted to
discussion. In Section~5, we summarize the results.

\section{Numerical methods to study global \\ dy\-na\-mics}

For detecting chaos in dynamical systems, variational methods and
methods of spectral analysis are mostly used. The essence of the
variational methods consists in an analysis of the time evolution
of trajectories with close initial conditions in phase space. In
the case of chaotic dynamics, the trajectories diverge with time
exponentially. A detailed analysis and a comparison of variational
methods and methods of spectral analysis can be found in
\cite{MDCG13}.

Numerical methods to study global behaviour of dynamical systems
include, primarily, techniques based on massive computations of
Lyapunov exponents \cite{MS98,SM03,PS13,LSS17,LSR18}, fast
Lyapunov indicators \cite{PLD02}, mean exponential growth number
\cite{CS00,CGS03}, fundamental frequencies of motion (frequency
analysis) \cite{CUM09,LC09}.

A classical method to determine the rate of divergence of close
trajectories in phase space is the method based on computing the
Lyapunov exponents \cite{O68,BGS76,BGGS80}. A dynamical
system with $N$ degrees of freedom has $2N$ Lyapunov exponents
(LE), but, in practice, only the maximum LE is usually determined.
By increasing the length of the time interval on which the maximum
LE is calculated, for a regular orbit the value of the numerically
determined finite-time maximum LE tends to zero, and for a chaotic
orbit it tends to some positive non-zero value. To obtain the full
Lyapunov spectrum, the HQRB-method (Householder QR-Based), in
particular, can be efficiently used, developed in \cite{BUP97}.
A comparison of various methods to compute LEs is given in
\cite{TSR01}. Note that even very long computations may often be
insufficient to distinguish between chaos and regular behaviour,
and to reveal the authentic LEs, because the computed LEs, are, in
fact, finite-time and local in nature; see discussion in
\cite{KLM18}.

The LE method is computationally expensive (see, e.g., a
discussion in \cite{SK02}). To reduce these costs, various
analogues (surrogates) of the Lyapunov exponents were developed.
The most popular among them are MEGNO \cite{CS00,CGS03} and FLI
\cite{FLF97}. The main idea of FLI, as proposed in
\cite{FLF97}, is to track the distance between two trajectories of
the phase space that are initially close to each other. If at some
stage of the integration the distance between the trajectories
exceeds a given critical value (the threshold criterion), the
dynamics is stated to be chaotic.

The MEGNO method was proposed in \cite{CS00, CGS03}. The MEGNO
parameter specifies the exponential growth factor of nearby
orbits, averaged in a particular way over a finite time interval.
In the case of a regular orbit, the value of the MEGNO parameter
is approximately constant; for a chaotic orbit, the value of the
MEGNO parameter increases with the length of the segment on which
the integration is performed. MEGNO and FLI allow one to identify
chaotic domains in phase space in much less (by 2--3 orders of
magnitude) integration times, in comparison with the LE method.
However, they do not provide any accurate estimates of the genuine
Lyapunov exponents; only local approximate estimates can be
obtained.

It should be noted that various simplifications and assumptions in
the LE surrogates may often lead to erroneous assessments of the
type of a trajectory. In particular, a disadvantage of the FLI
method consists in an ambiguity in the choice of the threshold
criterion for the identification of chaotic trajectories. Using
MEGNO may as well lead to ambiguous conclusions; as shown in
\cite{MDCG13}, in the case of a divided phase space, MEGNO may
characterize the regular component ambiguously. A software
package description for calculating various indicators of chaos
(including LE, FLI, and MEGNO) can be found in \cite{CMD14}.

A major spectral method is the method of frequency analysis (FA).
Its description and theoretical justification are given in
\cite{L90,L93,L03}. For regular orbits, the fundamental
frequencies are constant, while for chaotic orbits they are not
constantly defined, actions and angles varying randomly.
Performing the FA at separate time intervals, one can numerically
determine the current fundamental frequencies and find out whether
they vary when going from one time interval to another, i.e.,
determine the character of the dynamics. Examples of
implementation of the FA technique, as proposed in \cite{L90,L93}
in the form of a numerical analysis of fundamental frequencies,
can be found in \cite{VDQM10,VDQRW12}.

In addition to the general opportunity of identification of
regular and chaotic domains in phase space, FA allows one to
identify locations of resonances. However, FA is laborious; it may
require up to $\sim 30$\% of the computing time more than that
required by the LE method in one and the same problem
\cite{MDCG13}.

In celestial mechanics, to identify chaos in orbital or rotational
motion of celestial bodies, a number of specific methods were
proposed: the maximum eccentricity method (MEM) \cite{DPLS04,
SHE11}; methods based on massive numerical assessments of the
escape/encounter conditions \cite{HW99,PLFD03}; the reversibility
error method (REM) \cite{PGT17}. However, they are not
mathematically justified in any rigorous way. Moreover, the
criteria used in them for separating trajectories into chaotic and
regular ones are only approximate, similar to the case of FLI.

\section{The Poincar\'e recurrence method}

In this Section, we elaborate a general method, the
\textit{Poincar\'e recurrence method} (PRM), to study global
dynamics. The method is based on a massive numerical analysis of
Poincar\'e recurrences of orbits on fine grids of initial data
(or values of parameters).

\subsection{Basics of the PRM}

The notion of the Poincar\'e recurrence is of great methodological
value due to the existence of the famous {\it Poincar\'e
recurrence theorem} \cite{P1890}, valid in a broad class of
dynamical systems, including Hamiltonian systems on which we
concentrate here. Generally, the theorem states that for a
volume-conserving continuous one-to-one mapping $g$, transforming
a bounded domain $D$ of Euclidian space in itself ($g D = D$), in
any neighbourhood $U$ of any point of $D$ there exists a point $x$
that returns to $U$: $g^n x \in U$ at some $n$ (see
\cite{Arnold89}). In other words, any dynamical system of certain
type (in particular, with bounded phase space) recurs eventually,
though it may take much time, to any neighbourhood of its initial
state.

Although the theorem is valid for systems with bounded phase
space, the notion of Poincar\'e recurrence is defined for any
dynamical system. In particular, the PR method developed in this
article can be used, with minimal modifications, in systems with
non-bounded phase space, as demonstrated further on in
Subsection~\ref{PRM_3B}.

In various statistical applications, the so-called recurrence plot
technique and the recurrence quantification analysis, mostly
dealing with various data series, became more and more popular in
the last decades \cite{MRT07,M08}. The recurrence plot is defined
as a set of pairs of time instants when a dynamical system returns
to the same position in phase space. The recurrence plot technique
has been already used in celestial mechanics: in \cite{ABC04}, the
stability of selected exoplanetary systems was globally
characterized by the R\'enyi entropy, which was calculated by
using the recurrence plot technique. Computations of first
recurrence times were performed to construct a bifurcation diagram
for the standard map in \cite[figure~5]{MB13}.

And generally, computations of Poincar\'e recurrences have been
already broadly used in assessments of properties of dynamical
chaos in Hamiltonian systems. Various aspects of statistics of
Poincar\'e recurrences were numerically and analytically explored
in \cite{CS81,CS84,Sh10,C90,C99,CK08}.

The PRM, as proposed in this article, is intended for
construction of stability charts. A stability chart, in the sense
used here, is any global representation of the behaviour of any
parameters, characterizing instabilities (such parameters as LE,
FLI, MEGNO, or local diffusion rates) of a dynamical system, on a
two-dimensional plane of initial conditions or parameters of the
system. To produce a stability chart by means of PRM, the
Poincar\'e recurrences are computed on a uniform grid in the plane
of initial conditions of two selected variables (with all other
initial conditions and the system parameters fixed), or on a
uniform grid in the plane of two selected parameters (with all
other parameters and the initial conditions fixed). The grid is
defined in such a way that both regular and chaotic types of
trajectories can be analyzed on the subject of the properties of
their Poincar\'e recurrences in a representative way.

The Poincar\'e recurrences are calculated as follows. At each node
of the defined grid, a neighbourhood of the initial point of
motion of size $\varepsilon$ (either a sphere of radius
$\varepsilon$ or a box with size $\varepsilon$)
is defined in the phase space. By integrating numerically
equations of motion, a time instant $T_\mathrm{r}$ is fixed when
the trajectory returns to the given neighbourhood of the initial
point. The integration is over when either the first Poincar\'e
recurrence occurs or the end of the specified integration time
interval is over (thus, no Poincar\'e recurrence time is fixed).
Then, the durations of the recurrences are represented graphically
on the grid; say, in a colour grade.

Note that, in the current code, we define the recurrence box
either as a direct product of small linear intervals or as a small
sphere. Of course, other possibilities exist. However, as one may
expect (and this has been readily confirmed by our test numerical
experiments), this is not the shape of the recurrence box that is
important for the final clarity of a stability diagram, but its
size first of all.

Grids with various numbers of nodes can be used. A key point is
the choice of $\varepsilon$ at a given computation time; this
issue is discussed further on in Section~\ref{Discussion}.

To obtain a non-noisy stability chart, one should normally
choose the size of the recurrence box to be much smaller (by
orders of magnitude) than the length of the trajectory in one
recurrence. Therefore, any non-zero box-size corrections to the
computed length of a recurrence are ignored in the current version
of the codes. The middle point of the trajectory interval inside
the recurrence box \cite[Figure~1]{NSP12} might be a better
recurrence landmark otherwise.

In the following Subsections, PRM is featured separately for the
cases of bounded and non-bounded phase spaces, because the
embodiments of the PRM in each of the cases are specific. To
assess the performance of the PRM, we simultaneously use the
traditional LE method, and compare the results.

\subsection{The Poincar\'e recurrence method: the case of \\
bound\-ed phase space} \label{PRM_HH}

In the case of bounded phase space, the Poincar\'e recurrences are
counted with respect to a neighbourhood of a starting point in the
phase space. We take the H\'enon--Heiles system \cite{HH64} as a
paradigm for demonstrating the opportunities of PRM in this case.
It is in this problem that for the first time a chaotic behavior
was detected in Hamiltonian mechanics \cite{HH64}. The Hamiltonian
of the H\'enon--Heiles problem is given by

\begin{equation}
\label{hamilt} H = \frac{1}{2} \, (p_1^2 + p_2^2 + q_1^2 + q_2^2)
+ q_1^2 q_2 - \frac{1}{3} \, q_2^3 \, ,
\label{HHh}
\end{equation}

\noindent where $q_1$, $q_2$ are the canonical coordinates, and
$p_1$, $p_2$ are the conjugate canonical momenta.

Poincar\'e sections of the system's phase space were constructed
and domains of the chaotic motion were identified, using numerical
integration, in \cite{HH64}. With increasing the energy, the
chaotic domains grow in size, and at the energy value $E \equiv H
= 1/6$, practically all phase space of the possible motion is
chaotic \cite{SM03,HH64}. Note that the H\'enon--Heiles problem
was demonstrated \cite{FLF97} to be an example of the
effectiveness of the FLI method, in comparison with FA, in
detecting a chaotic behaviour.

Using a PRM\_HH code, described below, we compute the Poincar\'e
recurrences for a set of initial data defined on a uniform grid in
the plane ($p_2$,~$q_2$); the section is defined at $q_1 = 0$, and
$p_1$ are calculated by equation~(\ref{hamilt}) at $E = 0.1$. As
shown in \cite{SM03}, at $E = 0.1$ the chaotic domain takes
$\approx 20$\% of the whole phase space. Therefore, both regular
and chaotic types of trajectories can be analyzed on the subject
of the properties of their Poincar\'e recurrences in a
representative way.

The code, in accord with the general algorithm described
above, is organized as follows. At each node of the grid of
initial values (${p_i}_0$, ${q_i}_0 $), $i = 1,2$, a sphere of
radius $\varepsilon$ is defined in the phase space. By integrating
numerically the equations of motion specified by
Hamiltonian~(\ref{HHh}), the time instant $T_\mathrm{r}$ of
recurrence is fixed when the trajectory returns to the given
neighbourhood of the initial point, i.e., when
$\sum_{i=1,2}[(p_i-{p_i}_0)^2+(q_i-{q_i}_0)^2] \le \varepsilon^2$,
where $p_i$, $q_i$ are the current values of the canonical
variables.

Note that we have also tried ``box'' (brick-like) neighbourhoods,
of the same volume, in this problem. No effect on the final
results have been observed, as expected. (The box neighbourhoods
are also alternatively used in the next problem, considered in
Subsection~\ref{PRM_3B}.)

We use grids with various numbers of nodes: $100 \times 100$, $300
\times 300$ and $500 \times 500$. The intervals for the initial
$p_2$ and $q_2$ are defined as $p_2 \in [-0.5,0.5]$, $q_2 \in
[-0.4, 0.6]$. For a given energy value, all the trajectories in
the bounded phase space of Hamiltonian~(\ref{hamilt}) are
intersected by the defined subset of the ($p_2$, $q_2$) plane.

At $E = 0.08$, the fraction of chaos in the phase space is small
\cite{SM03}, and taking $\varepsilon = 10^{-2}$ provides the
Poincar\'e recurrence times $T_\mathrm{r} \le 10^3$ for 99\% of
the studied trajectories. On decreasing $\varepsilon$ to
$10^{-3}$, one has $T_\mathrm{r} \le 3 \times 10^4$ for 99\% of
the studied trajectories. If one takes $\varepsilon = 10^{-4}$,
then the integration time interval $t = 10^5$ turns out to be too
small to obtain any informative statistics on the distribution of
the Poincar\'e recurrences. For example, on the initial data grid
$100 \times 100$ the Poincar\'e recurrence times are fixed for
only 1\% of the studied trajectories. Therefore, to obtain the
results presented below, we have set $\varepsilon = 10^{-3}$ and
$t = 10^5$.

In addition to computing the Poincar\'e recurrences, the Lyapunov
times have been computed also, on the same grid of the initial
conditions and on the same time interval of integration. The
Lyapunov time is defined as $T_\mathrm{L} = 1/L$, where $L$ is the
maximum Lyapunov exponent (in fact, the maximum finite-time
local Lyapunov exponent). The calculation of the Lyapunov
exponents (finite-time local LE) in the H\'enon--Heiles
problem was carried out using the HQRB method in
\cite{BUP97,SK02}; for more details, see \cite{SM03}.

\begin{figure}[ht!]
\centering
\includegraphics[width=0.75\textwidth]{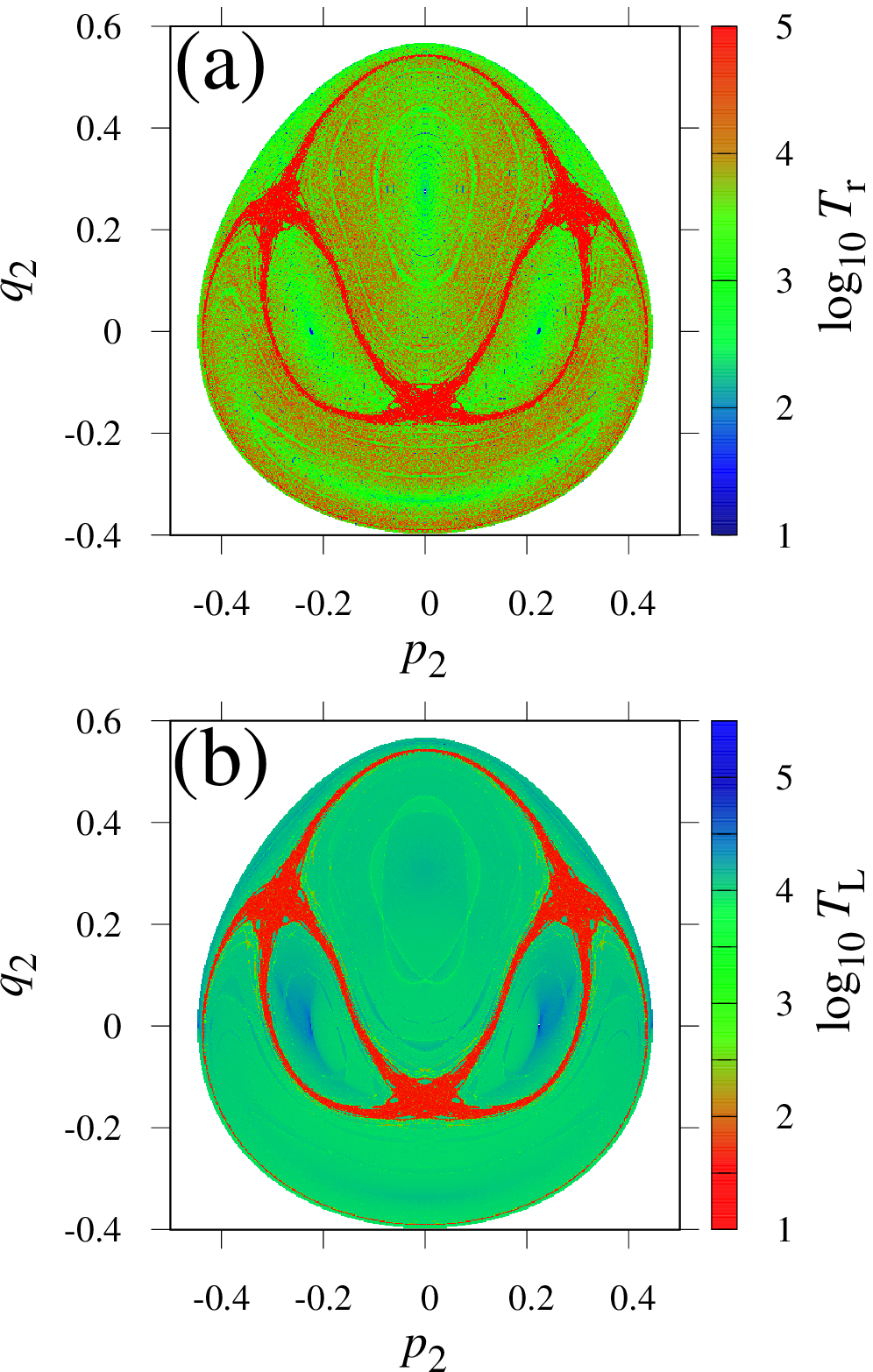}
\caption{(a)~The Poincar\'e recurrence chart for the
H\'enon--Heiles system, in the ($p_2$,~$q_2$) plane, at $E = 0.1$.
Red colour corresponds to $T_\mathrm{r} > 10^5$. (b)~The Lyapunov
time chart for the same system. Red colour corresponds to
$T_\mathrm{L} < 100$. In the both cases (a) and (b), the
integration time $T_\mathrm{int} = 10^5$.} \label{fig_HH_rec_time}
\end{figure}

Fig.~\ref{fig_HH_rec_time} shows the ($p_2$, $q_2$) diagrams with
the Poincar\'e recurrence and Lyapunov times indicated in a colour
grade. The diagrams allow one to judge on the structure of the
phase space of the H\'enon--Heiles system. The Poincar\'e
recurrence and Lyapunov times are calculated on a $500 \times 500$
grid with $\approx 150000$ nodes inside the bounded phase space.
In Fig.~\ref{fig_HH_rec_time}a, the Poincar\'e recurrence times
$T_\mathrm{r} \le 100$ (blue colour) correspond to the
trajectories passing through the centers of various resonances or
close to them. The Poincar\'e recurrence times $10^2 <
T_\mathrm{r} \le 2 \times 10^4$ (green colour) correspond to the
librational trajectories far from centers of resonances. The
Poincar\'e recurrence times $2 \times 10^4 < T_\mathrm{r} \le 5
\times 10^4$ (light red colour) correspond to the regular
trajectories located far from the resonances, and also probably to
weakly chaotic trajectories. The ring-like red areas all
correspond to chaotic trajectories; they have $T_\mathrm{r} >
10^5$.

To separate the trajectories into regular and chaotic ones, the
method proposed in \cite{MS98} is used. Its essence consists in
the analysis of the modal structure of the differential
distribution of the values of the computed Lyapunov exponents
(in fact, finite-time local Lyapunov exponents) computed on a
grid of initial data or values of parameters. Generally, the
distribution has two peaks, one fixed and one moving when the
computation time is increased. The fixed one corresponds to the
chaotic trajectories. On the contrary, the peak corresponding to
the regular trajectories moves along the horizontal axis towards
smaller computed finite-time LE values (towards larger
values of the Lyapunov times). Identifying the center of the gap
between the peaks, one obtains a numerical criterion for
separating the regular and chaotic trajectories.

In Fig.~\ref{fig_HH_rec_time}b, the Lyapunov times $T_\mathrm{L} <
100$ (red colour) correspond to the chaotic trajectories. Most of
the regular trajectories have $T_\mathrm{L}
> 10^4$ (blue colour). The trajectories with $10^3 < T_\mathrm{L} < 10^4$
(green colour) are also probably regular, but they are located
close to separatrices of resonances. Note the good structural
agreement between Fig.~\ref{fig_HH_rec_time}a and
Fig.~\ref{fig_HH_rec_time}b, regarding the locations and sizes of
the areas with the same character of dynamics.

\begin{figure}[ht!]
\centering
\includegraphics[width=0.7\textwidth]{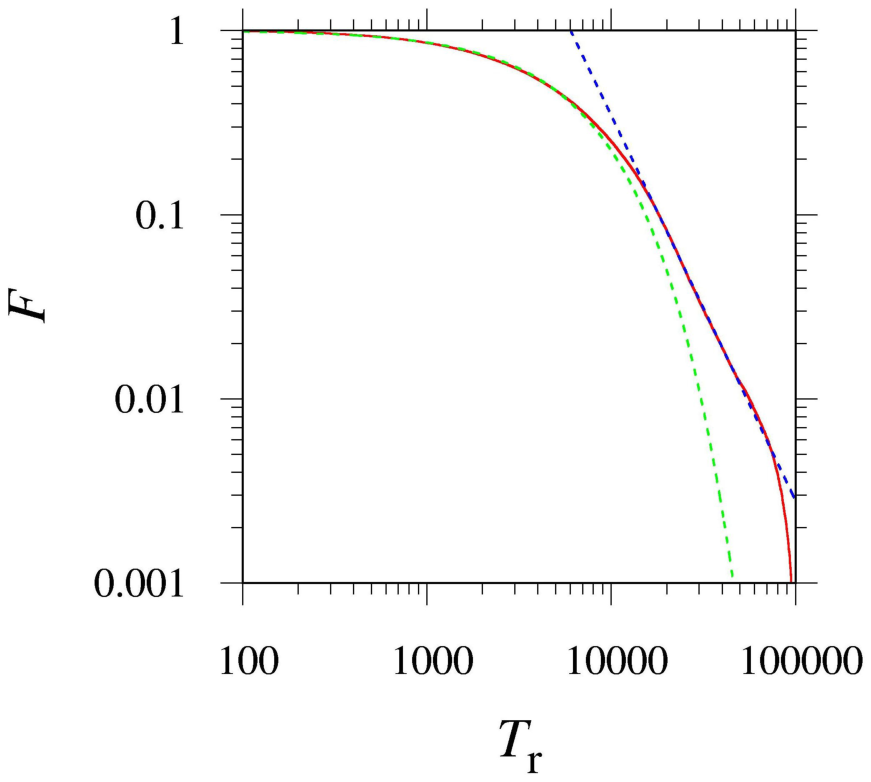}
\caption{The integral distribution of the Poincar\'e
recurrence times in the H\'enon--Heiles system (the solid curve).
The dashed curves represent fitting functions as explained in the
text. The parameters and initial data are the same as in
Fig.~\ref{fig_HH_rec_time}.}
\label{fig_HH_dist_rec_time}
\end{figure}

Fig.~\ref{fig_HH_dist_rec_time} shows the normalized integral
distribution of the Poincar\'e recurrence times. The distribution
subdivides into two parts: $T_\mathrm{r} \in [0, 10^4]$ and
$T_\mathrm{r} \in [2 \times 10^4, 8 \times 10^4]$. The first part
is naturally fitted by the exponential function $F \propto \exp
(-\alpha T_ \mathrm{r})$, where $\alpha = 1.5 \times 10^{-4}$
(green dashed curve), and the second part is naturally fitted by
the power-law function $F \propto {T_\mathrm{r}}^{-\beta}$, where
$\beta = 2.09$ (blue dashed straight line). In the both cases, the
correlation coefficient for the fitting function is ${\cal R} =
0.99$.

\subsection{Implementation of the PRM\_HH problem}
\label{subsecHHc}

The algorithm for calculating the Poincar\'e recurrence times for
the considered system with a bounded phase space (namely, the
H\'enon--Heiles system) is implemented in the PRM\_HH code,
written in Fortran90.\footnote{The code is available at
https://doi.org/10.5281/zenodo.3228905}
The total length of the code is only 180 lines (excluding the integrator).

To integrate the equations of motion, the Dormand--Prince
integrator DOP853 \cite{HNW93}, realizing the 8th order
Runge--Kutta method with an automatic time step size control, is
used. The maximum value of the time step size is set to $10^{-5}$,
and the local error tolerance to $10^{-12}$.

The program consists of the main part (PRM\_HH) and a subroutine
(calc\_rec\_time). In the calc\_rec\_time, the DOP853 integrator
is invoked, including subroutines fcn (where the equations of
motion are defined) and solout (where the recurrence condition is
checked after each integration step). Thus, in the PRM\_HH main
program loop, the subroutine calc\_rec\_time (t, t\_end, y) is
called, where y contains initial data for the integration.

\medskip

\noindent {\it INPUT.} The parameters and initial conditions for
the integration (the energy value, the initial data grid and the
radius of the neighborhood of the initial point where the
recurrence is fixed) are set directly in the PRM\_HH program body.
They are given by:

\begin{itemize}
\item H : the energy of the system; in the given problem, it takes
values within $[0,1 / 6] $. \item EPS : the radius of the
neighborhood (in which the Poincar\'e recurrence is fixed) of a
point in the phase space. \item T = 0 and T\_END  specify the
integration time interval. \item Q\_2\_INIT, Q\_2\_END,
P\_2\_INIT, and P\_2\_END define the borders of a uniform grid of
initial data in the plane ($p_2 $, $q_2 $). \item N\_GRID : the
number of steps along the axes $p_2$ and $q_2$.
\end{itemize}

\noindent {\it OUTPUT.} The output of the program PRM\_HH is
directed to the file \linebreak rec\_time.dat. The first and
second columns in the file contain the initial conditions in the
plane ($p_2$, $q_2$). The third column contains the recurrence
time rec\_time. If the integration time is too short to determine
the recurrence time, the value of the upper limit of the
integration time plus one is written to the third column.

Thus, at the end of the simulation, the PRM\_HH code gives
$N_{p_2} \times N_{q_2}=\mathrm{N\_GRID}\times\mathrm{N\_GRID}$
Poincar\'e recurrence time values.

\subsection{The Poincar\'e recurrence method: the case of \\
non-bounded phase space}
\label{PRM_3B}

In the case of non-bounded phase space, the Poincar\'e recurrences
are defined with respect to a neighbourhood of a starting point in
the phase space and with respect to the ``escape'' separatrix
(e.g., parabolic separatrix in the hierarchical restricted
three-body problem).

Let us consider, as a representative paradigm, the circumbinary
dynamics of a passively gravitating particle in the framework of
the restricted planar three-body problem. The mass parameter is
defined as $\mu = m_2/(m_1 + m_2)$, where $m_1 \ge m_2$ are the
masses of the primaries. The simulation is made in the synodic
reference frame. The Hamiltonian of the problem is given by

\begin{equation}
H=\frac{1}{2}\left(P_X^2+P_Y^2\right)+Y P_X-P_Y X-V(X,Y) ,
\label{H3b}
\end{equation}

\noindent where $X$, $Y$ are the Cartesian barycentric coordinates
of a passively gravitating tertiary and $P_X$, $P_Y$ are their
conjugate momenta, $V(X,Y)$ is the gravitational potential (see,
e.g., \cite{Morbi02,MD99}):

\begin{equation}
V(X,Y)=\frac{1-\mu}{R_1}+\frac{\mu}{R_2} ,
\end{equation}

\noindent where $R_1=\left[ \left( X-\mu \right)^2+Y^2
\right]^{1/2}$ and $R_2=\left[ \left( X+\left( 1-\mu \right)
\right)^2+Y^2 \right]^{1/2}$.

We use the integration code with the Levi--Civita regularization
(see \cite{Cel10,Cel02} for the equations). The number of the
tertiary's orbital revolutions serves to measure the first
recurrence times.

To assess the PRM performance, we use two methods in parallel: the
PRM and the LE methods. The PRM method is implemented in the
PRM\_3B code. The computation of orbits is based on a previous
code used to compute phase space fractal structure of the dynamics
governed by Hamiltonian~(\ref{H3b}) \cite{RLS16}. The integrator
used is the DOP853 integrator \cite{HNW93}, the same as described
above in Subsection~\ref{subsecHHc}.

In the PRM, the first recurrence is fixed when the following
conditions start to be satisfied:

\begin{equation}
X \in X_0 \pm \Delta X, \quad Y \in Y_0 \pm \Delta Y, \quad P_X
\in P_{X_0} \pm \Delta P_X, \quad P_Y \in P_{Y_0} \pm \Delta P_Y ,
\label{PR_conditions}
\end{equation}

\noindent where $X_0, Y_0, P_{X_0}, P_{Y_0}$ are the initial
conditions in the synodic reference frame, and $\Delta X = \Delta
Y = \Delta P_X = \Delta P_Y = 10^{-3}$. Thus, here the
$\varepsilon$ neighbourhood is defined as a box, instead of a
sphere used above in the H\'enon--Heiles problem
(Subsection~\ref{subsecHHc}).

We compute the dynamics on $201\times201$ grid nodes in the plane
``pericentric distance -- eccentricity'' ($q$--$e$) of the initial
conditions for the tertiary. The PRM\_3B code gives $201\times201
= 40401$ values of the number of the orbital revolutions of the
tertiary before the first recurrence.

The LE code computes the LE global charts. To compute the maximum
Lyapunov exponent (in fact, the maximum finite-time local
Lyapunov exponent), the code integrates the variational
equations, simultaneously with the equations of motion. The code
gives $201\times201 = 40401$ values of the maximum Lyapunov
exponent (the maximum finite-time local LE) after $T=10^{5}$
orbital revolutions of the binary.

\begin{figure}[ht!]
\centering
\includegraphics[width=0.6\textwidth]{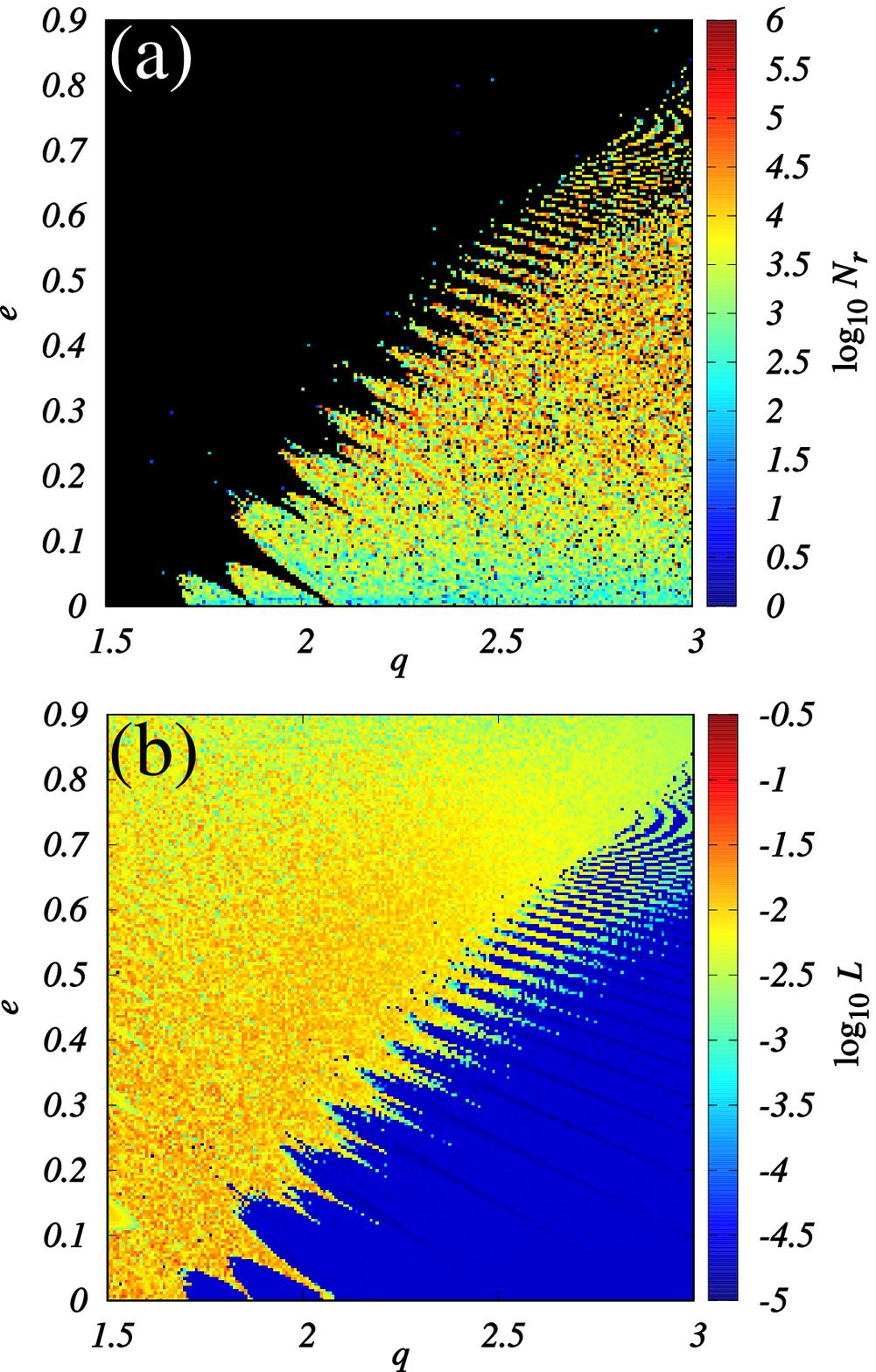}
\caption{(a)~The Poincar\'e recurrence chart in the ($q$,~$e$)
(``pericentric distance -- eccentricity'') plane in the planar
restricted three-body problem at $\mu = 0.1$. $N_r$ is the number
of the tertiary's orbital revolutions before the recurrence. The
integration time $T_\mathrm{int} = 10^6$. Black colour corresponds
to the orbits without recurrences within the integration time
$T_\mathrm{int}$. (b)~The Lyapunov time chart for the same
system.} \label{mu01}
\end{figure}

A comparison of the stability diagrams computed using the LE and
PRM is presented in Figure~\ref{mu01}. The choice of the
($q$,~$e$) (``pericentric distance -- eccentricity'') plane is
justified by the dynamical nature of the problem; this choice is
quite usual in problems concerning the circumbinary motion in
celestial-mechanical systems \cite{PS13}. The emergence of the
``teeth'' at the order/chaos boundary is due to the fractal
resonant structure of the border; the most prominent teeth are
formed by the overlap of subresonances of integer and half-integer
mean motion resonances between the particle and the central
gravitating binary (see \cite{PS13}). A close agreement between
the outcomes of application of the two methods is apparent.

\subsection{Implementation of the PRM\_3B problem}

The algorithm for calculating the Poincar\'e recurrence times for
a particular case of the unbounded phase space (namely, the
restricted three-body problem) is implemented in the PRM\_3B
program written in Fortran90.\footnote{The code is available at
https://doi.org/10.5281/zenodo.3228905}
The code length is $600$ lines (excluding the integrator).

Typically, the code makes a loop over $N_e$ initial values of the
eccentricity $e$ for a fixed initial pericentric distance $q$. A
Python code generates $N_q$ executables with different $q$.

The Levi--Civita regularization is employed to treat close
encounters of the bodies. For each integration step, the
regularization code invokes three changes of the reference frame.
It takes $150$ lines. To integrate the equations of motion, the
DOP853 integrator is used, the same as described above in
Subsection~\ref{subsecHHc}.

\medskip

\noindent {\it INPUT.} Several parameters are initially allocated
to the PRM\_3B code:

\begin{itemize}
 \item N : the number of the computed particle trajectories.
 \item MU : the mass parameter.
 \item RMAX : the maximum size (radius) of the orbits. If a particle were going
beyond this limit, it is regarded to have been ejected from
the system.
 \item RMIN1 : the minimum radius around the mass $m_1$. If a particle were
going below this limit, we assume a collision with $m_1$.
 \item RMIN2 : the same parameter as above, but for $m_2$.
 \item TMAX : the maximum time of simulation (counted in the binary's
orbital revolutions).
 \item TAU : the maximum allowed integration step size.
 \item EPS : the precision of the integration.
 \item EXMIN : the minimum initial eccentricity.
 \item EXMAX : the maximum initial eccentricity.
\end{itemize}

\noindent {\it OUTPUT.} At the end of the simulation, the PRM\_3B
code gives $N_r$ values of the number of the tertiary's orbital
revolutions before the first recurrence. The process is repeated
$N_q$ times on different computer cores.

\section{Poincar\'e recurrence statistics and diffusion rates}
\label{sec_stat}

The average time of recurrence (to one and the same subset of
phase space) can be roughly related in many cases (in particular,
in the case of the standard map \cite{C79}, describing an infinite
set of interacting resonances) to the diffusion rate by the
following formula

\begin{equation}
\tau \sim \frac{(\Delta y)^2}{D_y}
\label{dr_CS81}
\end{equation}

\noindent \cite[p.~11--12]{CS81}, where $\tau$ is the mean
recurrence time, $\Delta y$ is the characteristic distance in an
appropriate variable, $D_y$ is the diffusion rate in this
variable. In many-dimensional systems, a characteristic recurrence
time can be roughly estimated by identifying the ``slowest'' (that
exhibiting the slowest variation) variable in the system and, by
applying Equation~(\ref{dr_CS81}) for motion in this variable,
assessing $\tau$.

Therefore, the average return time can be set to be equal to the
average diffusion time. Formula~(\ref{dr_CS81}) can be
appropriate, as a simple but effective basic relation, in many
applications. As soon as PRM allows one to construct charts of the
diffusion timescales or rates, one can introduce a kind of the
``dynamical temperature'' \cite{S07PhysA} to characterize the
global dynamical behaviour of any system under study.

The character of the distribution of Poincar\'e recurrences at
large time\-scales is determined by the stickiness effect;
generally, the decay is algebraic \cite{CS81,CS84,Sh10}. Starting
with the pioneering work by Chirikov and Shepelyansky \cite{CS81},
the algebraic decay in the recurrence statistics in Hamiltonian
systems with divided phase space was considered, in particular, in
\cite{CS81,CS84,C90,C99,CK08,Sh10}. Chirikov~\cite{C90}, using his
resonant theory of critical phenomena in Hamiltonian dynamics,
predicted the critical exponent $\alpha$ in the integral
recurrence distribution

\begin{equation}
F(T_{\rm r}) \propto T_{\rm r}^{-\alpha}
\label{fta}
\end{equation}

\noindent to be equal to 3/2. (The integral distribution function
$F(T_{\rm r})$ is defined as the fraction of the recurrences that
have the duration greater than $T_{\rm r}$.)

In massive numerical simulations of dynamics of various
Hamiltonian systems, the algebraic decay was explored in
\cite{CK08}. System-dependent power-law exponents were revealed;
however, the ``universal'' average exponent turned out to be
well-defined: $\alpha = 1.57 \pm 0.03$ \cite{CK08}. This is quite
close to the theoretical $3/2$ value cited above. In celestial
mechanics, the algebraic decay was revealed in numerical
experiments on chaotic asteroidal dynamics \cite{SS96,SS97}. It
was found that the tail of the integral distribution of the time
intervals $T_{\rm r}$ between jumps of the eccentricity of
asteroids in the vicinity of the 3/1 mean-motion resonance with
Jupiter is algebraic:

\begin{equation}
F \propto T_{\rm r}^{-\alpha} ,
\end{equation}

\noindent where $\alpha \approx 1.5$--$1.7$.

Local properties of chaotic diffusion in Hamiltonian systems were
studied in \cite{RS96}; the statistics of exit times from
high-order resonances were explored in \cite{CS99}. In the both
studies, the results are in agreement with the Greene--MacKay
theory \cite{G79,McK83} of the critical golden curve. The longest
Poincar\'e recurrences were obtained in \cite{FS13}. As in
\cite{CK08}, it was concluded that the longest recurrences
originate from non-golden islands.

Thus, the power-law statistical relations between $T_\mathrm{L}$
and $T_\mathrm{r}$ are expected to emerge on long timescales, when
sticking of trajectories to chaos borders starts to dominate in
the statistics (see discussion in \cite{S98PLA}; also see figure~6
in \cite{MH97}, or figures~1 and 2 in \cite{S98PLA}). The
relationship between the recurrence times $T_{\rm r}$ and the
Lyapunov times $T_{\rm L}$ in systems with the stickiness effect
is generally quadratic \cite{S98PLA,S10PRE}. Note that the
long-term recurrence distributions, as well as relationships
between $T_{\rm r}$ and $T_{\rm L}$, in systems with non-bounded
phase space where escapes are possible, can have various power-law
indices, though their algebraic form is sustained \cite{S10PRE}.

In the current study, we have used relatively short computation
times --- short enough to fix most of the first recurrences on a
given grid. On much longer timescales, when sticking phenomena
come into play and recurrence statistics can be potentially
gathered for each node on the grid, comparisons between PRM and LE
charts, made in parallel, can be employed to establish and
massively study statistical $T_{\rm L}$--$T_{\rm r}$
relationships; that is why any application of the LE and PRM
methods in parallel can be of particular interest.

\begin{figure}[ht!]
\centering
\includegraphics[width=0.75\textwidth]{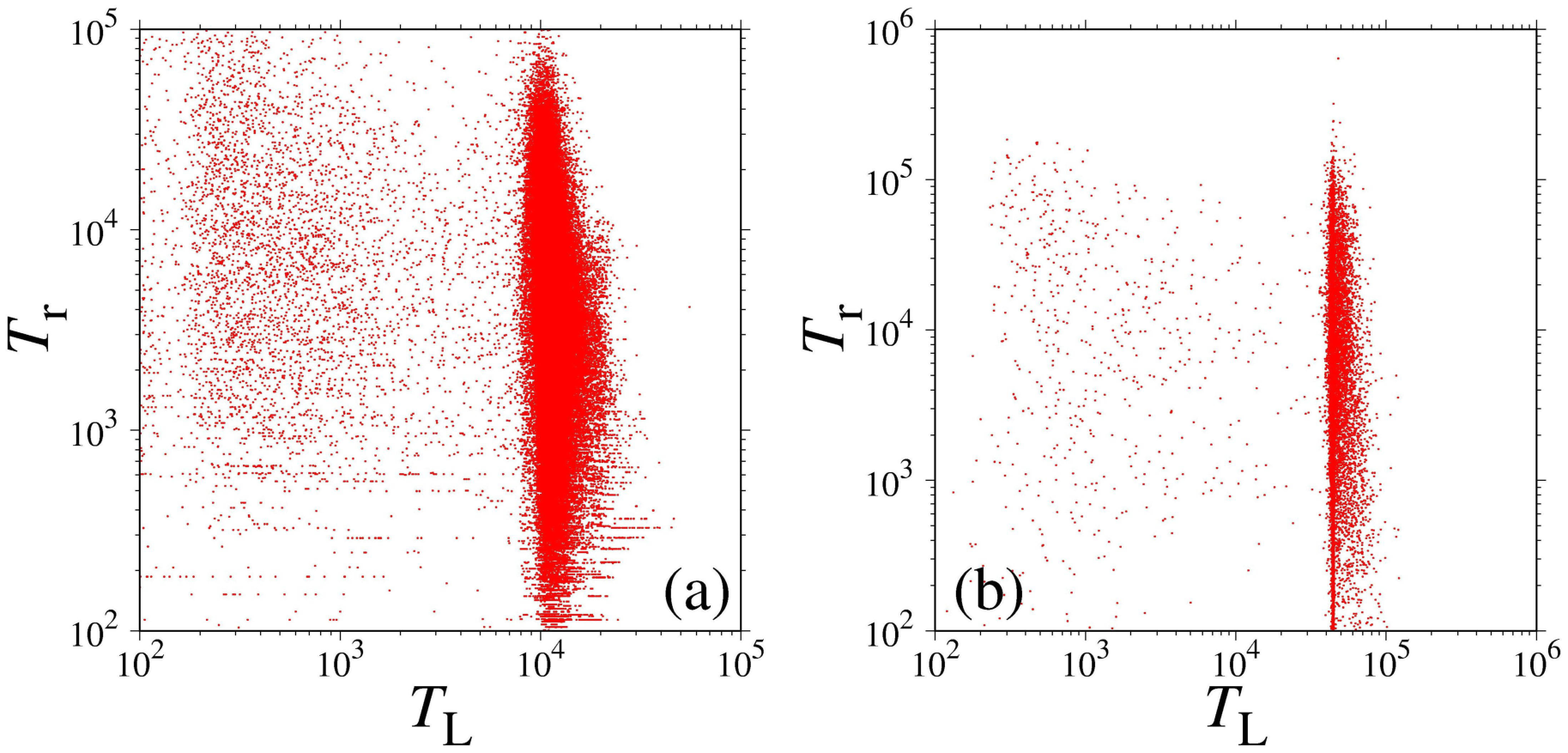}
\caption{The relationships between the Lyapunov and
Poincar\'e recurrence times in the considered systems: the
H\'enon--Heiles system (a) and the three-body problem (b).}
\label{fig_4}
\end{figure}

In Fig.~\ref{fig_4}, correlation plots
``$T_\mathrm{L}$--$T_\mathrm{r}$'' (Lyapunov time -- recurrence
time) for the both systems considered in this article are
presented. They have been constructed for the same data that were
computed to construct the PR charts, in such a way: for a
trajectory starting at a point in a PR chart, $T_\mathrm{L}$ is
fixed at the time $T_\mathrm{r}$ when the first recurrence is
fixed; the set of ($T_\mathrm{L}$,~$T_\mathrm{r}$) points over all
initial data forms the correlation plot. In fact, the both
``$T_\mathrm{L}$--$T_\mathrm{r}$'' plots show no correlation (only
broad scatter, apart from the vertical pile-ups corresponding to
regular trajectories), but one should not expect any
straightforward correlation between $T_\mathrm{L}$ and
$T_\mathrm{r}$ here, as soon as they are restricted by a
relatively small time limit of the computation. Besides, as
discussed above, the nature of emerging correlations, if any, can
be rather diverse and non-rigorous; see also \cite{MF96,BNP10}.
Therefore, PR charts have an independent value, in this sense:
they cannot be reproduced by any transformation of LE data.

As soon as relatively small time limits are set for the
computation of PR charts, sticking phenomena are relatively
unimportant. Besides, the measure of the critical component (where
the sticking occurs) of the phase space is also usually small (see
\cite{C90}); therefore, it does not affect the quality of PR
charts.

\section{Discussion}
\label{Discussion}

The assumption on the Euclidean metric to define the neighbourhood
where recurrences are fixed has been made throughout the article,
but other possibilities exist and they can be studied in the
course of the further development of the method.

Note that for the angle variables, the interval of variation
(normally $2 \pi$), in the current algorithm version is divided in
the same proportion as the intervals of variation of the momentum
variables with respect to the approximate size of the (bounded)
phase space in the momentum variables.

A key point in the both considered cases of bounded and unbounded
phase space is the choice of the size $\varepsilon$ of the
``recurrence-fixing sphere'' (i.e., the neighbourhood of the
initial point, where the recurrences are fixed) at a given
computation time; or, alternatively, a lower time limit for the
computation time, given the size $\varepsilon$.

For a bounded phase space, the lower time limit can be roughly
estimated assuming the approximate ergodicity of the chaotic
motion (excluding the critical component, which has low measure,
as noted above): $\Delta t / t_\mathrm{comp} \approx \Delta V /
V_\mathrm{s} \sim \varepsilon^n / V_\mathrm{s}$,  where $\Delta t$
is the time of residence of the trajectory in a point's
neighbourhood where recurrences are fixed, $t_\mathrm{comp}$ is
the time of computation, $V_\mathrm{s}$ is the full volume of the
phase space, $\Delta V \sim \varepsilon^n$ is the the volume of
the $\varepsilon$-box where recurrences are fixed, and $n$ is the
dimension of the phase space. Therefore, the minimum computation
time, allowing for at least a single expected recurrence, is $\sim
V_\mathrm{s} / \varepsilon^n$, in time steps of integration
(assuming the step is constant).

However, in practice, an appropriate time of computation is easily
evaluated empirically, by trying its higher and higher values
until the PR chart becomes noiseless.

Alternatively, an effective $\varepsilon$ can be easily evaluated
by fine-tuning the needed resolution of a dynamical chart of a
system under study. The algorithm is as follows: at first the size
of the recurrence-fixing sphere or box is chosen corresponding to
the desired resolution of a constructed stability diagram or phase
portrait; then, the computations are performed on a timescale of
at least one magnitude greater than the duration of the
first-fixed (over all the grid) recurrence. If this timescale
cannot be achieved using available computer resources, the desired
resolution should be decreased.

The novel PRM and the custom LE methods can be used in concert, so
that to utilize the best properties of both; when the computation
time is long enough for the sticking phenomenon to come into play,
such an approach would allow one to use statistical relationships
between the Lyapunov and diffusion timescales, when making
predictions for the long-term qualitative dynamical behaviour.

It should be underlined that the global charts of the massively
computed Poincar\'e recurrence times provide direct global
representations of spatial distributions of the local diffusion
times. The charts constructed in the quantities inverse to the
recurrence times provide massive measures of the local diffusion
rates, thus giving the picture of the global behaviour of the {\it
dynamical temperature} (defined analogously as in \cite{S07PhysA})
of any system under study.

Finally, note that for all examples provided in this article, the
integrator time step upper limit (set to $10^{-5}$ for PRM\_HH and
$2\pi\times10^{-3}$ for PRM\_3B) is small enough so that there is
no need, as established empirically, for any step diminishing
whenever the trajectory approaches a desired neighborhood.
Besides, in our computations, the local error tolerance of the
Dormand--Prince integrator was set to $10^{-12}$ for PRM\_HH (and
$10^{-10}$ for PRM\_3B), thus smaller than the chosen
$\varepsilon$ values by many orders of magnitude, therefore, the
integrator accuracy was by far sufficient. Due to the essential
sensitivity of the chaotic motion to the initial conditions,
nearby initial conditions may give rather different PR times.
However, on fine enough grids of initial data, the corresponding
``noise'' in the PR charts is suppressed, due to the statistical
averaging of the effect.

In this article, we have concentrated on the Hamiltonian
systems. Extensions of the method to the realm of dissipative
systems might be warranted; we leave this promising possibility
for a future work. Note also that the PRM can be developed further
on to incorporate calculations of second and consecutive
recurrences. This may favor to suppress any ``noisy'' appearances
in PRM diagrams. This opportunity is also left for a future
analysis.

\section{Conclusions}
\label{Concl}

We have shown that the novel PRM and the custom LE methods expose
the global dynamics almost identically, but the PRM allows one to
construct, in some approximation, charts of diffusion rates. This
ability reveals the major advantage of the novel method over the
custom global numerical tools (LE, FLI, MEGNO, FA). Moreover,
it is algorithmically simple and straightforward to apply.

\section*{Acknowledgements}

The authors express their gratitude to the referees for useful
remarks, which helped to improve the article. The authors are most
thankful to D.L.Shepe\-lyansky for valuable comments. I.I.S. and
A.V.M. were supported in part by the Russian Foundation for Basic
Research (project No.~17-02-00028) and by the Programme CP19--270
of Fundamental Research of the Russian Academy of Sciences. I.I.S.
was supported in part (in Sections~\ref{sec_stat} and
\ref{Discussion}) by the Russian Science Foundation (project
No.~19-12-11010). A.V.M. was supported in part by the Tomsk State
University competitiveness improvement programme. G.R. and J.L.
were supported by the French ``Investissements d'Ave\-nir''
program, project ISITE-BFC (contract ANR-15-IDEX-0003) and by the
Bourgogne Franche-Comt\'e Region 2017-2020 APEX project
(conventions 2017Y-06426, 2017Y-06413, 2017Y-07534).

\end{document}